\documentclass[twocolumn]{jpsj2} %% two-column layout
\title{Deformation of Electronic Structures Due to CoO$_6$ Distortion and Phase Diagrams of Na$_x$CoO$_2 \cdot y$H$_2$O}
\author
{Masahito {\sc Mochizuki}\footnote{E-mail address:
mochizuki@riken.jp} and Masao {\sc Ogata}$^1$}
\inst
{RIKEN, Hirosawa Wako, Saitama 351-0198, Japan \\
 $^1$Department of Physics, University of Tokyo, Hongo, Bunkyo-ku, Tokyo 113-0033, Japan}
\recdate{\today}

\abst{Motivated by recently reported experimental phase diagrams, we study the effects of CoO$_6$ distortion on the electronic structure in Na$_x$CoO$_2 \cdot y$H$_2$O. We construct the multiband tight-binding model by employing the LDA result. Analyzing this model, we show the deformation of band dispersions and Fermi-surface topology as functions of CoO$_2$-layer thickness. Considering these results together with previous theoretical ones, we propose a possible schematic phase diagram with three successive phases: the extended $s$-wave superconductivity (SC), the magnetic order, and the spin-triplet SC phases when the Co valence number $s$ is +3.4. A phase diagram with only one phase of spin-triplet SC is also proposed for the $s$=+3.5 case.}

\kword
{Co-oxide superconductor, multiorbital tight-binding model, phase diagram, Fermi-surface topology}

\begin{document}
\sloppy
\maketitle
Since the discovery of superconductivity (SC) in Na$_x$CoO$_2 \cdot y$H$_2$O,~\cite{Takada03,Takada05,Sakurai06} several experiments have suggested a spin-fluctuation-mediated unconventional pairing.~\cite{Ishida03,Fujimoto03,Kato03,Ihara04a,Ihara04b,Ihara06}
%$1/T_1$ measured by NQR experiments showed no coherence peak at $T_c$ and 
%$T^3$-behavior below $T_c$, which suggest non-$s$-wave pairing and line nodes 
%in the gap.~\cite{Ishida03,Fujimoto03,Kato03} 
%Heat capacity and $\mu$S~\cite{Higemoto04,Uemura04} measurements 
%have also suggested possible line nodes. 
%Ihara $et$ $al.$ showed that there is a correlation between $T_c$ and the 
%strength of magnetic fluctuation by NQR experiments.~\cite{Ihara04a,Ihara04b,Ihara06}
%They measured $1/T_1$ by preparing several samples and showed
%that a higher-$T_c$ sample shows a stronger enhancement of 
%$1/T_1$ at $T_c$. In addition to these experiments, various experiments 
%on the nonhydrate compounds have observed several characteristic 
%behaviors of strongly correlated electron systems 
%as well as magnetic orderings, which can also be circumstantial evidence for 
%unconventional pairing due to an electron-correlation mechanism.
%Thus, this material has attracted great interest. 
However, its SC properties have been controversial, probably because the sample preparation needs delicate control. Even its pairing symmetry has not been clarified yet. Several experiments have provided results that are inconsistent with each other, such as those for upper critical field $H_{c2}$~\cite{Takada05}, heat capacity~\cite{Oeschler05,Jin05} and phase diagrams~\cite{Chen04,Schaak05}. This makes the situation more complicated. In particular, phase diagrams proposed in the early stage of research are contradictory. One group proposed a constant $T_c$ as a function of Na content~\cite{Chen04}, while another group proposed a dome-shaped behavior of $T_c$~\cite{Schaak05}.

It has recently been pointed out that the valence of Co ions in this material is not +3.65 as naively expected from the Na content $x$ of $\sim$0.35.~\cite{Milne04,Karppinen04} Instead, the actual valence is +3.4-3.5 irrespective of $x$ because of the presence of oxonium ions (H$_3$O$^+$). Here, the valence $s$ is directly related to the number of $t_{2g}$ electrons per Co ion ($n_{t_{2g}}$) as $n_{t_{2g}}=9-s$. Thus, the tuning of $x$ does not necessarily correspond to the carrier doping. This fact has not been recognized so far, and the contradiction of experimentally obtained phase diagrams mentioned above is, in fact, attributed to this point.

Elaborate phase diagrams have been experimentally obtained by precisely determining the Co valence $s$ by Sakurai $et$ $al$.~\cite{Sakurai06,Sakurai05a,Sakurai05b} They proposed a phase diagram in the plane of temperature and $x$. They found that when one prepares samples whose $s$ value is constant at +3.4, a SC phase (SC1), a magnetically ordered phase (MO) and another SC phase (SC2) successively appear as $x$ increases. On the other hand, for samples with $s=$+3.5, only one SC phase appears. Here, we can consider that $x$ controls the trigonal CoO$_6$ distortion along the $c$-axis (in other words, the thickness of the CoO$_2$ layers, which is constructed from edge-shared CoO$_6$ octahedra): A larger-$x$ sample has thicker CoO$_2$ layers. Several groups have shown the relationship between $T_c$ and the crystallographic $c$-axis parameter $c/a$ and that between $T_c$ and the thickness of CoO$_2$ layers.~\cite{Lynn03,Sakurai04}. In addition, similar phase diagrams have also been proposed recently by other groups.~\cite{Ihara04b,Ihara06,Sato06,Michioka06,Zheng06b}

These phase diagrams indicate that the electronic structures in this material are quite sensitive to the thickness of CoO$_2$ layers. Therefore, in order to clarify the properties and mechanism of the SC as well as the magnetic properties, it is important to consider the effects of CoO$_6$ distortion more seriously. In fact, the present authors previously examined these effects by tuning the trigonal crystal field (CF) parameter in the multiorbital Hubbard model.~\cite{Mochizuki05} It was shown that the increasing trigonal CoO$_6$ distortion pushes the $e'_g$ bands up, resulting in the formation of hole pockets near the K-points, and thus the ferromagnetic and triplet-pairing instabilities are enhanced. However, in this study, the changes in hopping parameters due to the distortion were not taken into account, and Sakurai's phase diagram with three successive phases (SC1, MO and SC2 phases) was not reproduced.

In this letter, we study the effects of CoO$_6$ distortion on the electronic structure in Na$_x$CoO$_2 \cdot y$H$_2$O by constructing a multiband tight-binding model including not only the fivefold Co $3d$ orbitals but also the threefold O $2p$ orbitals. (Note that our previous model contains only the threefold Co $t_{2g}$ orbitals.~\cite{Mochizuki05}) Using this model, we first study microscopically the changes in both band dispersions and Fermi surface (FS) topology. We show that when $s$ is fixed at +3.4, three types of FSs successively appear as the CoO$_6$ distortion increases. We discuss that this FS deformation can explain well Sakurai's phase diagrams for $s=$+3.4. In particular, we predict that two SC phases in this phase diagram have different pairing states, one is a spin-singlet and the other is a spin-triplet. We also study the band-structure deformation and the phase diagram for the $s=$+3.5 case. In addition, the contradictory results of $H_{c2}$~\cite{Takada05} and heat capacity~\cite{Oeschler05,Jin05} measurements as well as those for the nature of magnetic correlations are also discussed.

%%%%%%%%%%%%%%%%%%%%%%%%%%%%%%%%%%%%%%%%%%%%%%%%%%%%%%%%%%%%%
\begin{figure}[tdp]
\includegraphics[scale=0.32]{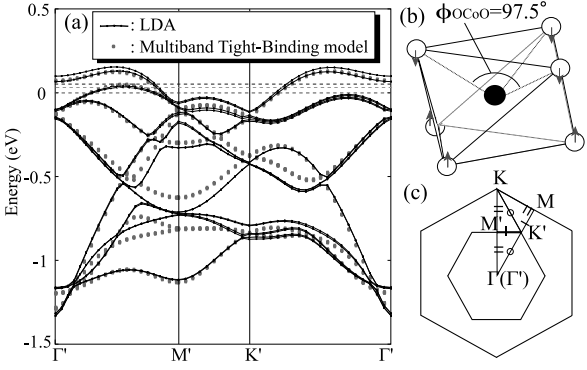}
%\epsfile{file=fig1.eps,scale=0.32}
\caption{(a) Co $t_{2g}$ band dispersion calculated from multiband tight-binding model for $\phi_{\rm OCoO}$=97.5$^{\circ}$ (dots), and that from LDA calculation for bilayer-hydrate system (solid lines).~\cite{Johannes04} Three horizontal dashed lines denote the Fermi levels for $s=$+3.4, +3.5 and +3.67 cases, respectively, from upper to lower. (b) Trigonal CoO$_6$ distortion along the $c$-axis. Arrows indicate the directions of O-ion shifts. (c) Original Brillouin zone and folded Brillouin zone for expanded unit cell (see text).}
\label{band1}
\end{figure}
%%%%%%%%%%%%%%%%%%%%%%%%%%%%%%%%%%%%%%%%%%%%%%%%%%%%%%%%%%%%%
The multiband tight-binding Hamiltonian is given by
%%%%%%%%%%%%%%%%%%%%%%%%%%%%%%%%%%%%%%%%%%%%%%%%%%%%%%%%%%%%%%%
\begin{eqnarray}
H^{dp}&=&H_{\rm cry.}
      +\sum_{i,\gamma,\sigma} \varepsilon_{d}
        d_{i\gamma\sigma}^{\dagger} d_{i\gamma\sigma}
      +\sum_{j,l,\sigma} \varepsilon_{p}
        p_{jl\sigma}^{\dagger} p_{jl\sigma} \nonumber \\
      &+&\sum_{i,\gamma,i',\gamma',\sigma} 
        t_{i\gamma,i'\gamma'}^{dd} d_{i\gamma\sigma}^{\dagger} 
                           d_{i'\gamma'\sigma}
      +\sum_{i,\gamma,j,l,\sigma} 
        t_{i\gamma,jl}^{dp}d_{i\gamma\sigma}^{\dagger} 
                            p_{jl\sigma}    \nonumber \\
      &+&\sum_{j,l,j',l',\sigma} 
        t_{jl,j'l'}^{pp}p_{jl\sigma}^{\dagger} 
                         p_{j'l'\sigma},
\label{dph}
\end{eqnarray}
%%%%%%%%%%%%%%%%%%%%%%%%%%%%%%%%%%%%%%%%%%%%%%%%%%%%%%%%%%%%%%%
where $d^{\dagger}$ and $p^{\dagger}$ are the creation operators of the Co $3d$ electron and the O $2p$ electron, respectively. The indices $i$ and $j$ run over the Co and O sites, respectively. The indices $\gamma$ and $l$ run over the fivefold $3d$ and threefold $2p$ orbitals, respectively. The first term represents CFs.
%due to the Coulomb repulsion between Co $3d$ and O $2p$ electrons. 
The second and third terms express the level energies of the Co 3$d$ and O 2$p$ orbitals, respectively. The other three terms express the $d$-$d$, $d$-$p$ and $p$-$p$ electron transfers, respectively. 

Here, the term $H_{\rm cry.}$ consists of two components:
%%%%%%%%%%%%%%%%%%%%%%%%%%%%%%%%%%%%%%%%%%%%%%%%%%%%%%%%%%%%%%%
\begin{equation}
H_{\rm cry.}=\sum_{i,m,\sigma} \Delta_{e_g}
           d_{im\sigma}^{\dagger} d_{im\sigma}
        +\sum_{i,m',m'',\sigma} D_{m'm''}
           d_{im'\sigma}^{\dagger} d_{im''\sigma},
\label{cry}
\end{equation}
%%%%%%%%%%%%%%%%%%%%%%%%%%%%%%%%%%%%%%%%%%%%%%%%%%%%%%%%%%%%%%%
where $m$ in the first term runs over the twofold Co $e_g$ orbitals, and $m'$ and $m''$ in the second term are indices of Co $t_{2g}$ orbitals. The first term represents the cubic CF generated by the octahedrally coordinated O ions, which contributes to the $t_{2g}$-$e_g$ splitting. The second term represents the trigonal CF due to the trigonal CoO$_6$ distortion where $D_{m'm''}=\frac{1}{3}\Delta_{\rm D3d}(\delta_{m'm''}-1)$, which contributes to the $a_{1g}$-$e'_g$ splitting of threefold $t_{2g}$ degeneracy.

The transfer integrals $t_{dd}$, $t_{dp}$ and $t_{pp}$ are determined by the Slater-Koster (SK) parameters:~\cite{Harrison89} $(dd\sigma)=-0.395$, $(dd\pi)=0.103$ and $(dd\delta)=0.026$ for the nearest-neighbor (NN) Co-Co bonds, $(pd\sigma)=-2.060$ and $(pd\pi)=0.989$ for the NN O-Co bonds, $(pp\sigma)_1=0.623$ and $(pp\pi)_1=-0.164$ for the NN O-O bonds, $(pp\sigma)_2=0.467$ and $(pp\pi)_2=-0.123$ for the second-NN O-O bonds, $(pp\sigma)_3=0.327$ and $(pp\pi)_3=-0.078$ for the third-NN O-O bonds, and $(pp\sigma)_4=0.101$ and $(pp\pi)_4=-0.024$ for the fourth- and fifth-NN O-O bonds. In addition, the charge-transfer energy $\varepsilon_{d}-\varepsilon_{p}$ is 2.20, and the CF parameters $\Delta_{e_g}$ and $\Delta_{\rm D3d}$ are 0.55 and 0.0, respectively. These parameter values are deduced by fitting the LDA result for bilayer-hydrate material.~\cite{Johannes04} The energy unit is eV.

The CoO$_2$ layer becomes thinner when the octahedra are more significantly contracted along the $c$-axis. We simulate this trigonal distortion by displacing the O ions along the trigonal directions (see  Fig.~\ref{band1}(b)). With this distortion, the OCoO bond angle ($\phi_{\rm OCoO}$) is no longer 90$^{\circ}$. The angle is increased further from 90$^{\circ}$ when the distortion is more significant. When we carry out the above fitting, $\phi_{\rm OCoO}$ is fixed at 97.5$^{\circ}$ along with the condition for which the LDA calculation was performed.~\cite{Johannes04,Jorgensen03}

In Fig.~\ref{band1}, we display the Co $t_{2g}$ band dispersions calculated using these parameters together with those obtained in the LDA calculation, which shows good coincidence. We take the expanded hexagonal unit cell with lattice vectors of length $\sqrt{3}a$, which results in a folded Brillouin zone (BZ) with one-third of the volume of the original one. The $\Gamma'$, ${\rm K}'$ and ${\rm M}'$ points are symmetry points defined on the folded BZ, and the relationship between the original and folded BZs is shown in Fig.~\ref{band1}(c).

From the relationship between $(pp\sigma)_1$ and $(pp\sigma)_2$, we see that $(pp\sigma)$ and $(pp\pi)$ for adjacent O ions are proportional to $l_{\rm OO}^{-2.19}$, where $l_{\rm OO}$ is the distance between two adjacent O ions. In addition, from the expression of the interatomic matrix element, we expect that $(pd\sigma)$ and $(pd\pi)$ are proportional to $l_{\rm CoO}^{-3.5}$, where $l_{\rm CoO}$ is the neighboring Co-O distance.~\cite{Harrison89} Using these relationships as well as the obtained SK parameters, we investigate the changes in band structure due to CoO$_6$ distortion by varying $\phi_{\rm OCoO}$.

Before discussing the results, we note that the trigonal CF parameter $\Delta_{\rm D3d}$ necessarily changes with varying $\phi_{\rm OCoO}$. Although the value monotonically increases with increasing $\phi_{\rm OCoO}$, its estimation is quite difficult and contains ambiguity because we need to estimate the value of the local dielectric constant for the Coulomb interaction between electrons on the O and Co ions. Thus, we first discuss the effects of transfer-integral variation due to the distortion by fixing $\Delta_{\rm D3d}$ to be constant, and then we discuss the effects of $\Delta_{\rm D3d}$ variation by assuming several ways of monotonic change.

%%%%%%%%%%%%%%%%%%%%%%%%%%%%%%%%%%%%%%%%%%%%%%%%%%%%%%%%%%%%
\begin{figure}[tdp]
\includegraphics[scale=0.35]{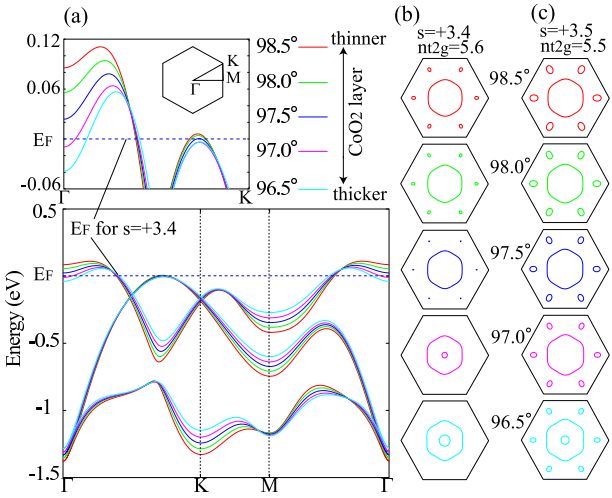}
%\epsfile{file=fig2.eps,scale=0.35}
\caption{(a) Band dispersions calculated from multiband tight-binding model for several $\phi_{\rm OCoO}$ values. Deformation of Fermi surface for $s$=+3.4 (b) and that for $s$=+3.5 (c) with varying $\phi_{\rm OCoO}$.}
\label{band2}
\end{figure}
%%%%%%%%%%%%%%%%%%%%%%%%%%%%%%%%%%%%%%%%%%%%%%%%%%%%%%%%%%%%%
Figure~\ref{band2}(a) shows the Co $t_{2g}$ band dispersions for several $\phi_{\rm OCoO}$ values. Here, $\Delta_{\rm D3d}$ is fixed to be constant at zero. The horizontal dashed line denotes the Fermi level for $s=$+3.4. As shown in this figure, both the $a_{1g}$ band at the $\Gamma$ point and the top of the $e'_g$ bands near the K points are located below the Fermi level with thick CoO$_2$ layers of $\phi_{\rm OCoO}$=96.5$^{\circ}$ and 97.0$^{\circ}$. As the CoO$_2$ becomes thinner, both are pushed up. Thus, both appear above the Fermi level when the CoO$_2$ layers are thin (see the cases of $\phi_{\rm OCoO}$=98.0$^{\circ}$ and 98.5$^{\circ}$).

We show the deformation of FS topology for both $s=$+3.4 and $s=$+3.5 cases in Figs.~\ref{band2}(b) and ~\ref{band2}(c), respectively. We see that in the case of $s=$+3.4, FS with two $a_{1g}$-band cylinders (FS1) appears for thick CoO$_2$ layers, while FS with one $a_{1g}$-band cylinder and six $e'_g$-band pockets (FS2) appears for thin CoO$_2$ layers. On the other hand, in the case of $s=$+3.5, FS1 does not appear even for a thick CoO$_2$ layer. Instead, for a wide range of $\phi_{\rm OCoO}$ values, FS2 appears, while another type of FS with two $a_{1g}$ cylinders and six $e'_g$ pockets (FS3) is realized when the layer is very thick with $\phi_{\rm OCoO}$=96.5$^{\circ}$.

%%%%%%%%%%%%%%%%%%%%%%%%%%%%%%%%%%%%%%%%%%%%%%%%%%%%%%%%%%%%
\begin{figure}[tdp]
\includegraphics[scale=0.4]{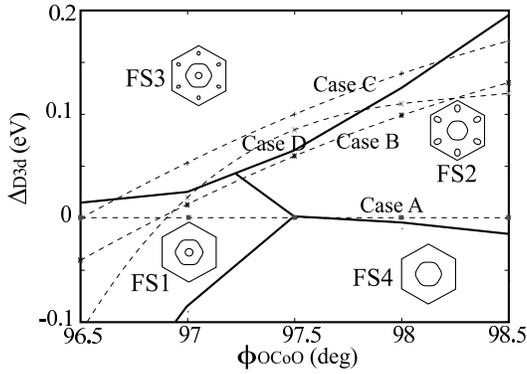}
%\epsfile{file=fig3.eps,scale=0.4}
\caption{Phase diagram of Fermi-surface topology for $s$=+3.4 in plane of $\phi_{\rm OCoO}$ and $\Delta_{\rm D3d}$.}
\label{fspd}
\end{figure}
%%%%%%%%%%%%%%%%%%%%%%%%%%%%%%%%%%%%%%%%%%%%%%%%%%%%%%%%%%%%%
Next, we discuss the effects of trigonal CF, i.e., $\Delta_{\rm D3d}$. In Fig.~\ref{fspd}, we show a phase diagram of FS topology in the plane of $\phi_{\rm OCoO}$ and $\Delta_{\rm D3d}$ for $s$=+3.4. We find four types of FSs: FS1, FS2, FS3 and FS4. Here, FS4 consists of only one $a_{1g}$ cylinder. In the actual material, we expect that $\Delta_{\rm D3d}$ will monotonically increase with $\phi_{\rm OCoO}$. Here, we consider several ways of increasing $\Delta_{\rm D3d}$, for example, as represented by lines for cases A, B, C and D in this figure. In the prior calculation, $\Delta_{\rm D3d}$ is constantly fixed at 0.0, which corresponds to the line denoted as case A. If $\Delta_{\rm D3d}$ increases along the line of case B, two types of FSs (FS1 and FS2) successively appear with increasing $\phi_{\rm OCoO}$. On the other hand, we see that three types of FSs, FS1, FS3 and FS2, successively appear for case C. For case D, in particular, the successive deformation of these three FSs occurs even within a tiny $\phi_{\rm OCoO}$ change of 1$^{\circ}$.

This phase diagram suggests that FS1 tends to be realized in a system with thick CoO$_2$ layers, while a system with thin CoO$_2$ layers tends to have FS2. In fact, the former is the FS which Kuroki $et$ $al.$ employed to propose a possible extended $s$-wave pairing,~\cite{Kuroki06} while the latter is the one on which several theoretical proposals of spin-triplet $p$- and $f$-wave pairings are based.~\cite{Kuroki04,Mochizuki04,Yanase04} As for the spin fluctuation, finite-momentum ($q\ne$0) fluctuation is predicted for FS1,~\cite{Kuroki06} while nearly ferromagnetic fluctuation at zero momentum ($q=$0) is predicted for FS2.~\cite{Kuroki04,Mochizuki04,Yanase04} The $q\ne$0 fluctuation for FS1 is enhanced by the electron-electron scattering between inner and outer $a_{1g}$ FSs owing to the intraorbital Coulomb interaction,~\cite{Kuroki06} while the $q=$0 fluctuation for FS2 is caused by the interband scattering between $a_{1g}$ and $e'_g$ bands owing to the Hund's-rule coupling.~\cite{Mochizuki04}

For a moderate thickness of the CoO$_2$ layers such as $\phi_{\rm OCoO}$=97.5$^{\circ}$, the FS topology strongly depends on $\Delta_{\rm D3d}$. If $\Delta_{\rm D3d}$ is small, as in case A, FS4 is realized. With such a FS, the system has no specific enhanced spin fluctuation, as discussed in previous work.~\cite{Mochizuki05} On the other hand, when $\Delta_{\rm D3d}$ is moderate or large, FS2 or FS3 appears as in cases B, C and D. With these FSs, we expect enhanced spin fluctuations.~\cite{Mochizuki05,Mochizuki04} In particular, we expect that the spin fluctuation is markedly enhanced for FS3 because both electron scattering between the inner and outer $a_{1g}$ cylinders and scattering between $a_{1g}$ cylinders and $e'_g$ pockets cooperatively contribute to the enhancement of spin fluctuation.

%%%%%%%%%%%%%%%%%%%%%%%%%%%%%%%%%%%%%%%%%%%%%%%%%%%%%%%%%%%%
\begin{figure}[tdp]
\includegraphics[scale=0.35]{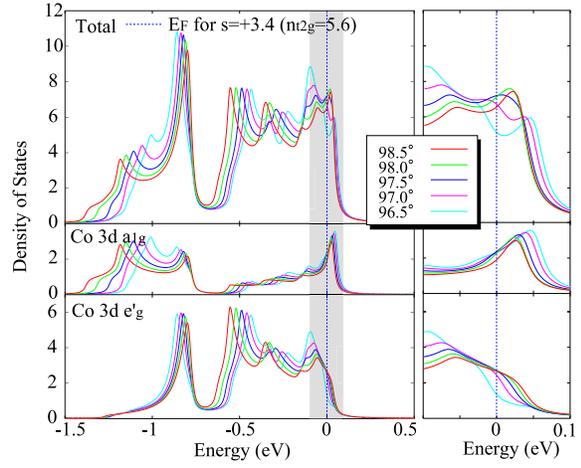}
%\epsfile{file=fig4.eps,scale=0.35}
\caption{Density of states (DOS) for case C. The left panel shows the total DOS, the partial DOS for the $a_{1g}$ orbital and that for the $e'_g$ orbital from upper to lower. The vertical dashed line denotes the Fermi level for $s$=+3.4. The right panel magnifies the shaded area in the left panel.}
\label{dos}
\end{figure}
%%%%%%%%%%%%%%%%%%%%%%%%%%%%%%%%%%%%%%%%%%%%%%%%%%%%%%%%%%%%%
In Fig.~\ref{dos}, we display the total density of states (DOS), the partial DOS for the Co $a_{1g}$ orbital ($a_{1g}$ DOS) and that for the Co $e'_g$ orbital ($e'_g$ DOS) for case C. The blue dashed line denotes the Fermi level for $s$=+3.4. This figure shows the following three aspects: [1] For thick CoO$_2$ layers (such as $\phi_{\rm OCoO}$=96.5$^{\circ}$), the $a_{1g}$ DOS is rather large while the $e'_g$ DOS is small. [2] For thin CoO$_2$ layers (such as $\phi_{\rm OCoO}$=98.5$^{\circ}$), the $e'_g$ DOS is rather large while the $a_{1g}$ DOS is small. [3] Both $a_{1g}$ and $e'_g$ DOSs are large for moderate thicknesses such as $\phi_{\rm OCoO}$=97.5$^{\circ}$ and 98.0$^{\circ}$. The situation in the moderate case is favorable for the enhancement of magnetic fluctuation due both to the Hund's-rule coupling between the $a_{1g}$ and $e'_g$ bands and to the intraorbital Coulomb interaction in the $a_{1g}$ band.

%%%%%%%%%%%%%%%%%%%%%%%%%%%%%%%%%%%%%%%%%%%%%%%%%%%%%%%%%%%%
\begin{figure}[tdp]
\includegraphics[scale=0.4]{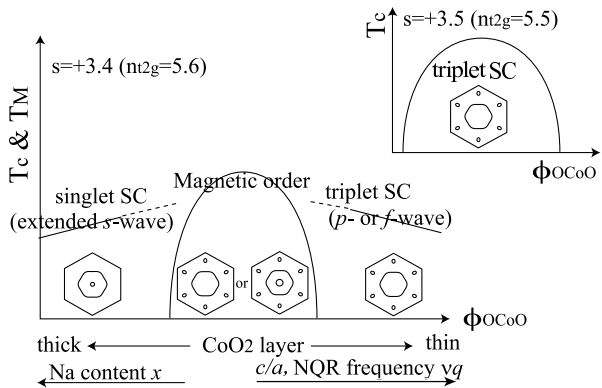}
%\epsfile{file=fig5.eps,scale=0.4}
\caption{Schematic phase diagram for the $s=$+3.4 case in plane of $\phi_{\rm OCoO}$ and temperature. The inset shows a schematic phase diagram for the $s=$+3.5 case.}
\label{scpd}
\end{figure}
%%%%%%%%%%%%%%%%%%%%%%%%%%%%%%%%%%%%%%%%%%%%%%%%%%%%%%%%%%%%%
Now, we discuss the experimentally obtained phase diagrams for $s$=+3.4 and $s$=+3.5 (see Fig.~\ref{scpd}) on the basis of the above results as well as previously obtained ones.~\cite{Kuroki06,Kuroki04,Mochizuki04,Yanase04} For $s$=+3.4, the observed two SC phases are likely to be the singlet extended $s$-wave pairing with FS1~\cite{Kuroki06} and the triplet $f$- or $p$-wave pairing with FS2~\cite{Kuroki04,Mochizuki04,Yanase04}, respectively. The inbetween MO phase is caused by the cooperative contributions from enhanced $q\ne0$ and $q=0$ spin fluctuations. This is owing both to the FS topology of FS3 (or FS2) and to the DOS structure in which both $a_{1g}$ and $e'_g$ DOSs are rather large. On the other hand, the experimental phase diagram for $s$=+3.5 consists of only one SC phase. The pairing state in this SC phase is expected to be the triplet $p$- or $f$-wave one with FS2, from Fig.~\ref{band2}(c). Note that the $c$-axis parameter $c/a$, NQR frequency $\nu_q$ and Na content $x$ appearing in Fig.~\ref{scpd} are scaled with the thickness of CoO$_2$ layers.~\cite{Ihara04b,Sakurai05a,Sakurai05b,Sakurai04,Zheng06b}

We speculate that NMR experiments, which recently observed the decreasing Knight shift below $T_c$, were performed for samples in the singlet-pairing phase.~\cite{Kobayashi03a,Kobayashi05,Zheng06,Yada06} We expect that a constant behavior may be observed in a sample with triplet pairing having thinner CoO$_2$ layers. It is also worth pointing out that the confusing results of $H_{c2}$~\cite{Takada05} and heat capacity~\cite{Oeschler05,Jin05} measurements may be attributed to the difference in pairing state depending on the sample. In addition, the contradictions of NMR/NQR and neutron-scattering results for the magnetic-correlation type may also be due to this sample dependence.~\cite{Ishida03,Fujimoto03,Kato03,Kobayashi03a,Ning04,Moyoshi06} Thus, it is highly desired to perform these experiments by determining where the sample is located in the phase diagram. Angle-resolved photoemission spectroscopy for the bilayer-hydrate system is also desired to detect the predicted FS deformation, since few measurements have been reported to date.~\cite{Shimojima06} 

To summarize, we have constructed the multiband tight-binding model including the full degeneracies of Co $3d$ and O $2p$ orbitals for Na$_x$CoO$_2 \cdot y$H$_2$O. Using this model, we have studied the effects of CoO$_6$ distortion on band dispersion, FS topology and DOS. On the basis of this analysis, we have proposed possible interpretations of experimental phase diagrams for both $s=$+3.4 and $s=$+3.5 cases. We have predicted that the following three phases successively appear in the $s=$+3.4 phase diagram with decreasing CoO$_2$-layer thickness: a SC phase with singlet extended $s$-wave pairing, a MO phase and another SC phase with triplet $p$- or $f$-wave pairing. On the other hand, only one SC phase with triplet pairing has been predicted for the $s=$+3.5 phase diagram. 

We thank H. Sakurai, K. Ishida, Y. Yanase, K. Kuroki, R. Arita, K. Yoshimura, G.-q. Zheng, Y. Kobayashi, and M. Sato for valuable discussions. This work is supported by a Grant-in-Aid for Scientific Research from MEXT and the RIKEN Special Postdoctoral Researcher Program.
%%%%%%

\end{document}